\documentclass[12pt]{article}

\usepackage{amsmath}
\usepackage{graphicx}

\newcommand{\ket}[1]{| #1 \rangle}
\newcommand{\bra}[1]{\langle #1 |}
\newcommand{\hcs}[1]{#1^\dagger #1}
\newcommand{\expv}[1]{\langle #1 \rangle}

\begin{document}

\begin{center}
{\Large\bf Information, fidelity, and reversibility
in general quantum measurements}
\vskip .6 cm
Hiroaki Terashima
\vskip .4 cm
{\it Department of Physics, Faculty of Education, Gunma University, \\
Maebashi, Gunma 371-8510, Japan}
\vskip .6 cm
\end{center}

\begin{abstract}
We present the amounts of
information, fidelity, and reversibility obtained by
arbitrary quantum measurements on completely unknown states.
These quantities are expressed as functions of
the singular values of a measurement operator
corresponding to the obtained outcome.
As an example, we consider a class of quantum measurements
with highly degenerate singular values
to discuss tradeoffs among
information, fidelity, and reversibility.
The tradeoffs are at the level of a single outcome,
in the sense that
the quantities pertain to each single outcome
rather than the average over all possible outcomes.
\end{abstract}

\begin{flushleft}
{\footnotesize
{\bf PACS}: 03.65.Ta, 03.67.-a\\
{\bf Keywords}: quantum measurement, quantum information
}
\end{flushleft}

\section{Introduction}
In quantum theory,
information about a physical system cannot be obtained
without affecting it
because quantum measurement inevitably changes
the state of the system via nonunitary state reduction.
This property of quantum measurement is profoundly interesting
for the foundations of quantum mechanics
and is of practical importance in quantum information processing
and communication~\cite{NieChu00},
such as in quantum cryptography~\cite{BenBra84,Ekert91,Bennet92,BeBrMe92}.
Therefore, the subject of a tradeoff
between information gain and state change has been discussed
by many authors~\cite{FucPer96,Banasz01,FucJac01,BanDev01,Barnum02,%
DArian03,Ozawa04,Maccon06,Sacchi06,BusSac06,Banasz06,BuHaHo07,FGNZ15}
over several years using various formulations.
For example, Banaszek~\cite{Banasz01} showed
an inequality between two fidelities quantifying
information gain and state change
and Ozawa~\cite{Ozawa04} generalized
Heisenberg's uncertainty relation for noise and disturbance
in quantum measurements.

On the other hand,
state change due to quantum measurement has been shown
not to be necessarily irreversible~\cite{UedKit92,UeImNa96,Ueda97}
if the measurement preserves all the information about the system,
though it was once widely believed to be irreversible such
that one could not recover the premeasurement state
from the postmeasurement one~\cite{LanLif77}.
In fact, in a physically reversible measurement~\cite{UeImNa96,Ueda97},
the premeasurement state can be recovered from
the postmeasurement one with a nonzero probability
of success via a second measurement,
called a reversing measurement.
Reversible measurements have been proposed
for various physical systems~\cite{Imamog93,Royer94,TerUed05,KorJor06,%
TerUed07,SuAlZu09,XuZho10}
and have been experimentally demonstrated
using a superconducting phase qubit~\cite{KNABHL08}
and a photonic qubit~\cite{KCRK09}.

Thus, it is natural to discuss
not only the size of the state change but also its reversibility
while considering the costs of information gain.
Intuitively, as measurements provide more information about a system,
one would expect that more information would result in
more change of a system's state along with reduced reversibility.
Moreover, whenever the reversing measurement
recovers the premeasurement state of the first measurement,
it erases all the information
obtained by the first measurement
(see the Erratum of Ref.~\cite{Royer94}).
In a different type of reversible measurement,
known as unitarily reversible measurement~\cite{MabZol96,NieCav97},
the premeasurement state can be recovered from the postmeasurement one
with unit probability via a unitary operation
although the measurement provides no information about the system.
Therefore, there are some tradeoffs among information gain, state change,
and physical reversibility in quantum measurement.

Such tradeoffs have been studied
in photodetection processes~\cite{Terash10} and
in single-qubit measurements~\cite{Terash11}.
These tradeoffs are at the level of a single outcome,
in contrast to conventional
ones~\cite{FucPer96,Banasz01,BanDev01,Barnum02,Sacchi06,Banasz06};
that is to say that the quantities affected
are those pertaining to each single outcome,
rather than those averaged over all possible outcomes.
This characteristic is desirable
for studying state recovery with information erasure
in a physically reversible measurement,
because it occurs not on average
but only when the reversing measurement
yields a preferred single outcome.
On the other hand, using quantities averaged over outcomes,
Cheong and Lee~\cite{CheLee12} demonstrated that
a tradeoff exists between information gain and physical reversibility,
which has been experimentally verified~\cite{CZXTLX14,LRHLK14}
using single photons.

In this paper,
we present the general formulas for information gain, state change,
and physical reversibility for an arbitrary quantum measurement
on a $d$-level system in a completely unknown state.
These formulas are more general versions of
those for an arbitrary quantum measurement
on a two-level system~\cite{Terash11}
and those for a projective measurement
on a $d$-level system~\cite{Terash11b}.
We present the evaluation of the amount of information gain
by the decrease in Shannon entropy~\cite{DArian03,TerUed07b},
the degree of state change by the fidelity~\cite{Uhlman76},
and the degree of physical reversibility by
the maximum successful probability of
the reversing measurement~\cite{KoaUed99}.
The formulas are written using the singular values of a measurement operator
corresponding to the outcome of the measurement.
Unfortunately, when some singular values are degenerate,
the formula for information gain is not useful
for numerical calculations due to apparent divergences.
Therefore, for the information gain,
we show another formula that is free from apparent divergences,
even when the singular values are degenerate.

The rest of this paper is organized as follows:
Section~\ref{sec:formulation} explains the procedure for quantifying
information gain, state change, and physical reversibility
and shows their explicit formulas.
Section~\ref{sec:degeneracy} deals with the degeneracy of singular values.
Section~\ref{sec:example} considers
a class of quantum measurements
with highly degenerate singular values
and discusses the tradeoffs among information gain, state change,
and physical reversibility.
Section~\ref{sec:conclude} summarizes our results.

\section{\label{sec:formulation}Formulation}

\subsection{Information gain}
We first consider the amount of information provided
by a quantum measurement.
To evaluate this amount, it is first assumed that
the premeasurement state of a system to be measured
is known to be one of a set of
predefined pure states $\{\ket{\psi(a)}\}$, $a=1,\ldots,N$,
each of which has an equal probability of $p(a)=1/N$,
although the index $a$ of the premeasurement state is unknown.
The lack of information about the state is then given by
\begin{equation}
  H_0=-\sum_a p(a)\log_2 p(a)=\log_2 N
\label{eq:H0}
\end{equation}
prior to measurement, where
the Shannon entropy has been used as a measure of uncertainty
rather than the von Neumann entropy of the mixed state
$\hat{\rho}=\sum_a p(a) \ket{\psi(a)}\bra{\psi(a)}$
because the uncertain information is the classical variable $a$
rather than the predefined quantum state $\ket{\psi(a)}$.
Each state $\ket{\psi(a)}$ can be expanded
in an orthonormal basis $\{\ket{i}\}$ as
\begin{equation}
  \ket{\psi(a)}=\sum_i c_i(a)\, \ket{i},
\label{eq:basis}
\end{equation}
where $i=1,2,\ldots,d$, and $d$ is the dimension of 
the Hilbert space associated with the system.
For the state to be normalized,
the coefficients $\{c_i(a)\}$ must satisfy
the normalization condition
\begin{equation}
\sum_i \left| c_i(a)\right|^2=1.
\label{eq:normalization}
\end{equation}
Since, in quantum measurements,
the system to be measured is usually in a completely unknown state,
the predefined states $\{\ket{\psi(a)}\}$ are assumed to be
all of the possible pure states of the system with $N\to\infty$.

A quantum measurement of the system can then be made
to obtain information about the state.
In general,
a quantum measurement is described by a set of
measurement operators $\{\hat{M}_m\}$~\cite{DavLew70,NieChu00}
that satisfy
\begin{equation}
\sum_m\hcs{\hat{M}_m}=\hat{I},
\label{eq:completeness}
\end{equation}
where $m$ denotes the outcome of the measurement
and $\hat{I}$ is the identity operator.
When the system is in a state $\ket{\psi}$,
the measurement $\{\hat{M}_m\}$ yields an outcome $m$ with probability
\begin{equation}
 p_m=\bra{\psi}\hcs{\hat{M}_m}\ket{\psi},
\label{eq:probability}
\end{equation}
changing the state into
\begin{equation}
\ket{\psi_m}=\frac{1}{\sqrt{p_m}}\,\hat{M}_m\ket{\psi}.
\label{eq:reduction}
\end{equation}
Here it has been assumed that the quantum measurement
is efficient~\cite{FucJac01} or ideal~\cite{NieCav97}
in the sense that the postmeasurement state is pure
if the premeasurement state is pure,
in order to focus on the quantum nature of measurement
by ignoring classical noise.
Each measurement operator $\hat{M}_m$
can be decomposed by singular-value decomposition as
\begin{equation}
 \hat{M}_m=\hat{U}_m\hat{D}_m\hat{V}_m,
\label{eq:decomposition}
\end{equation}
where $\hat{U}_m$ and $\hat{V}_m$ are unitary operators,
and $\hat{D}_m$ is a diagonal operator
in the orthonormal basis $\{\ket{i}\}$:
\begin{equation}
\hat{D}_m=\sum_{i} \lambda_{mi} \ket{i}\bra{i}.
\end{equation}
The diagonal elements $\{\lambda_{mi}\}$,
called the singular values of $\hat{M}_m$,
are not less than $0$ by definition and
are not greater than $1$ on the basis of Eq.~(\ref{eq:completeness}); that is,
\begin{equation}
  0\le\lambda_{mi}\le1
\end{equation}
for $i=1,2,\ldots,d$.
In this situation, where the measurement is performed on
one of all possible pure states $\{\ket{\psi(a)}\}$,
the unitary operator $\hat{V}_m$ can be removed
from the measurement operator given in Eq.~(\ref{eq:decomposition}) as
\begin{equation}
 \hat{M}_m=\hat{U}_m\hat{D}_m
\end{equation}
by relabeling the index $a$ as
$\ket{\psi'(a)}=\hat{V}_m\ket{\psi(a)}$ without loss of generality.
Furthermore,
the unitary operator $\hat{U}_m$ is irrelevant to information gain,
since the probability given by Eq.~(\ref{eq:probability})
is unaffected by $\hat{U}_m$.
Although it changes the state of the system as in Eq.~(\ref{eq:reduction}),
the state change caused by $\hat{U}_m$ can be recovered with unit probability
and no information loss
after the measurement by applying $\hat{U}_m^\dagger$ to the system.
Thus, to see the inevitable state change and irreversibility
caused by the extraction of information,
it suffices to set
the measurement operator of Eq.~(\ref{eq:decomposition}) equal to
\begin{equation}
 \hat{M}_m=\hat{D}_m.
\label{eq:operator}
\end{equation}

By substituting Eqs.~(\ref{eq:basis}) and (\ref{eq:operator})
into Eq.~(\ref{eq:probability}),
it is evident that the measurement yields
outcome $m$ with probability
\begin{equation}
  p(m|a)=\sum_{i} \lambda_{mi}^2\left| c_i(a)\right|^2
        \equiv q_m(a)
\label{eq:pma}
\end{equation}
when the premeasurement state of the system is $\ket{\psi(a)}$.
Since the probability of $\ket{\psi(a)}$ is $p(a)=1/N$,
the total probability of the outcome $m$ is given by
\begin{equation}
  p(m) =\sum_a  p(m|a)\,p(a)=\frac{1}{N}\sum_a q_m(a)=
     \overline{q_m},
\label{eq:pm}
\end{equation}
where the overline denotes the average over $a$,
\begin{equation}
   \overline{f} \equiv \frac{1}{N}\sum_a f(a).
\label{eq:overline}
\end{equation}
On the contrary, given the outcome $m$,
the probability of the premeasurement state $\ket{\psi(a)}$
can be calculated to be
\begin{equation}
  p(a|m) =\frac{p(m|a)\,p(a)}{p(m)}=\frac{q_m(a)}{N\,\overline{q_m}}
\label{eq:pam}
\end{equation}
according to Bayes' rule.
Therefore, after the measurement yields the outcome $m$,
the lack of information about the premeasurement state
decreases to the Shannon entropy
\begin{equation}
  H(m) =-\sum_a p(a|m)\log_2 p(a|m).
\end{equation}
Using this decrease in Shannon entropy~\cite{DArian03,TerUed07b},
the information provided by the measurement
with the outcome $m$ can be expressed as
\begin{equation}
  I(m) \equiv H_0-H(m)
    =\frac{\overline{q_m\log_2 q_m} -\overline{q_m}\log_2 \overline{q_m}}
     {\overline{q_m}},
\label{eq:Im}
\end{equation}
which is always positive and evidently free from
the divergent term $\log_2 N\to\infty$ in Eq.~(\ref{eq:H0}),
due to the assumption that $p(a)$ is uniform.
This quantity can be viewed as
the relative entropy (or the Kullback--Leibler divergence)~\cite{NieChu00}
of $p(a|m)$ to the uniform distribution $p(a)=1/N$,
\begin{equation}
  I(m)=\sum_a p(a|m)\, \log_2\frac{p(a|m)}{p(a)}.
\end{equation}

To explicitly calculate the information in Eq.~(\ref{eq:Im}),
it is necessary to average $q_m(a)$ and $q_m(a)\log_2 q_m(a)$ over
all possible pure states of the system, $\{\ket{\psi(a)}\}$.
As shown in Appendix \ref{sec:average},
a straightforward calculation gives
\begin{equation}
   \overline{q_m} = \frac{1}{d}\,\sigma_m^2,
\label{eq:qmbar}
\end{equation}
where $\sigma_m$ is the Hilbert--Schmidt norm of $\hat{M}_m$,
\begin{equation}
 \sigma_m = \sqrt{\textrm{Tr}\left(\hat{M}_m^\dagger\hat{M}_m\right)}
   =\sqrt{\sum_{i}\lambda_{mi}^2}.
\label{eq:normHS}
\end{equation}
On the other hand, it would be difficult to directly calculate
the average of $q_m(a)\log_2 q_m(a)$
using the method described in Appendix \ref{sec:average}.
However, in different contexts,
similar calculations have been performed
in various ways~\cite{Wootte90,Jones91,JoRoWo94}.
By applying the integral formula derived in Ref.~\cite{Jones91}
to this case, the following expression can be obtained:
\begin{equation}
\overline{q_m\log_2 q_m} =\frac{1}{d}\sum_{i}
  \frac{\lambda_{mi}^{2d}\log_2\lambda_{mi}^2}
       {\prod_{k\neq i}\left( \lambda_{mi}^2-\lambda_{mk}^2\right)}
   - \frac{1}{d\ln2}\Bigl[\eta(d)- 1\Bigr]\sigma_m^2,
\label{eq:qmlogqmbar}
\end{equation}
where $\eta(n)$ is defined by
\begin{equation}
 \eta(n) \equiv\sum^{n}_{k=1}\frac{1}{k}
     =1+\frac{1}{2}+\cdots+\frac{1}{n}.
\end{equation}
Note that in order to obtain the form of Eq.~(\ref{eq:qmlogqmbar})
from the integral formula,
it is necessary to use the identity
\begin{equation}
 \sum_{i}\frac{\lambda_{mi}^{2d}}
    {\prod_{k\neq i}\left( \lambda_{mi}^2-\lambda_{mk}^2\right)}
  =\sigma_m^2
\label{eq:identity4lambda}
\end{equation}
and the recurrence formula of the digamma function $\psi(z)$,
$\psi(z+1)= \psi(z)+1/z$.
By substituting Eqs.~(\ref{eq:qmbar}) and (\ref{eq:qmlogqmbar})
into Eq.~(\ref{eq:Im}),
the information can finally be expressed as
\begin{equation}
 I(m) =\log_2d-
         \frac{1}{\ln2}\Bigl[\eta(d)- 1\Bigr]
    -\log_2\sigma_m^2+\frac{1}{\sigma_m^2}
    \sum_{i}\frac{\lambda_{mi}^{2d}\log_2\lambda_{mi}^2}
      {\prod_{k\neq i}\left( \lambda_{mi}^2-\lambda_{mk}^2\right)}.
\label{eq:information}
\end{equation}
This function is invariant
under the interchange of any pair of singular values,
\begin{equation}
 \lambda_{mi} \longleftrightarrow \lambda_{mj}
 \quad \text{for any $(i,j)$},
\label{eq:interchange}
\end{equation}
as well as under the rescaling of all singular values
by a constant factor $c$,
\begin{equation}
\left(\lambda_{m1},\lambda_{m2},\ldots,\lambda_{md}\right) \to 
\left(c\lambda_{m1},c\lambda_{m2},\ldots,c\lambda_{md}\right),
\label{eq:rescale}
\end{equation}
because of Eq.~(\ref{eq:identity4lambda}).
If the singular values are normalized by
the rescaling factor of Eq.~(\ref{eq:rescale}) to $\sigma_m^2=1$,
the $\{\lambda_{mi}\}$-dependent part of Eq.~(\ref{eq:information}),
\begin{equation}
 Q=\log_2\sigma_m^2-\frac{1}{\sigma_m^2}\sum_{i}
  \frac{\lambda_{mi}^{2d}\log_2\lambda_{mi}^2}
         {\prod_{k\neq i}\left( \lambda_{mi}^2-\lambda_{mk}^2\right)},
\label{eq:subentropy}
\end{equation}
resembles the subentropy discussed in Ref.~\cite{JoRoWo94}.
However, these quantities have different meanings,
since the subentropy is a function
of the eigenvalues of the premeasurement density operator
$\hat{\rho}=\sum_a p(a) \ket{\psi(a)}\bra{\psi(a)}$,
rather than a function of the singular values
of the measurement operator $\hat{M}_m$.
For a fixed $d$, Eq.~(\ref{eq:subentropy}) satisfies
the inequality~\cite{JoRoWo94}
\begin{equation}
  0 \le Q \le \log_2d-
         \frac{1}{\ln2}\Bigl[\eta(d)- 1\Bigr].
\end{equation}
The lower bound is achieved when only one singular value is nonzero,
as in the projective measurement of rank $1$,
whereas the upper bound is achieved when all singular values are equal,
as in the identity operation.

The information in Eq.~(\ref{eq:Im})
is at the level of a single outcome in the sense that it has its value
when a single outcome $m$ has been obtained.
If $I(m)$ is averaged over all outcomes
with probabilities given by Eq.~(\ref{eq:pm}),
the mutual information~\cite{NieChu00}
of the random variables $\{a\}$ and $\{m\}$ is obtained:
\begin{equation}
   I\equiv\sum_m p(m)\,I(m)=
   \sum_{m,a} p(m,a)\, \log_2\frac{p(m,a)}{p(m)\,p(a)},
\label{eq:Iav}
\end{equation}
where $p(m,a)=p(m|a)\,p(a)$.
However, this is the amount of information
that is expected to be obtained on average
before the measurement, rather than the actual information $I(m)$.
While the average information expressed by Eq.~(\ref{eq:Iav})
is not discussed further in this paper,
the explicit form of $I$ is presented herein,
since it cannot be found in the literature.
It becomes
\begin{align}
 I &=\log_2d-
         \frac{1}{\ln2}\Bigl[\eta(d)- 1\Bigr] \notag \\
  &\qquad
    {}-\frac{1}{d}\sum_m\left[ \sigma_m^2\log_2\sigma_m^2
    -\sum_{i}\frac{\lambda_{mi}^{2d}\log_2\lambda_{mi}^2}
      {\prod_{k\neq i}\left( \lambda_{mi}^2-\lambda_{mk}^2\right)}\right]
\end{align}
from Eqs.~(\ref{eq:pm}), (\ref{eq:qmbar}), and (\ref{eq:information}),
with an identity resulting from
the trace of Eq.~(\ref{eq:completeness}),
\begin{equation}
   \sum_m \sigma_m^2=d.
\label{eq:identity4normHS}
\end{equation}

\subsection{State change}
Now the degree of state change caused by the measurement
as a cost of the information gain is considered.
When the premeasurement state of the system is $\ket{\psi(a)}$,
a measurement with outcome $m$ changes it to
\begin{equation}
   \ket{\psi(m,a)} = \frac{1}{\sqrt{q_m(a)}}\,\hat{D}_m
       \ket{\psi(a)}
\end{equation}
according to Eq.~(\ref{eq:reduction})
with Eqs.~(\ref{eq:operator}) and (\ref{eq:pma}).
This state change can be evaluated
using the fidelity~\cite{Uhlman76,NieChu00} as
\begin{equation}
   F(m,a) = \bigl|\expv{\psi(a)|\psi(m,a)}\bigr|
          =\frac{1}{\sqrt{q_m(a)}}
       \sum_{i}\lambda_{mi}\left| c_i(a)\right|^2
     \equiv \frac{f_m(a)}{\sqrt{q_m(a)}},
\label{eq:fma}
\end{equation}
which decreases as
the measurement changes the state of the system by a greater extent.
By averaging over the premeasurement states $\{\ket{\psi(a)}\}$
with probabilities given by Eq.~(\ref{eq:pam}),
the fidelity after the measurement with the outcome $m$ can be expressed as
\begin{equation}
 F(m) =\sum_a p(a|m)\bigl[F(m,a)\bigr]^2
     = \frac{\overline{f_m^2}}{\;\overline{q_m}\;},
\label{eq:Fm}
\end{equation}
where the squared fidelity,
rather than the fidelity, has been averaged for simplicity.

To explicitly calculate the fidelity in Eq.~(\ref{eq:Fm}),
it is necessary to average $\left[f_m(a)\right]^2$ over
all possible pure states of the system, $\{\ket{\psi(a)}\}$.
As shown in Appendix \ref{sec:average},
the average is given by
\begin{equation}
   \overline{f_m^2} =
    \frac{1}{d(d+1)}\left(\sigma_m^2+\tau_m^2\right),
\label{eq:fm2bar}
\end{equation}
where $\tau_m$ is the trace norm of $\hat{M}_m$,
\begin{equation}
  \tau_m = \textrm{Tr}\sqrt{\hat{M}_m^\dagger\hat{M}_m}
   =\sum_{i}\lambda_{mi}.
\label{eq:normT}
\end{equation}
By substituting Eqs.~(\ref{eq:qmbar}) and (\ref{eq:fm2bar})
into Eq.~(\ref{eq:Fm}), the fidelity can be obtained as follows:
\begin{equation}
    F(m) =\frac{1}{d+1}\left(\frac{\sigma_m^2
                +\tau_m^2}{\sigma_m^2}\right).
\label{eq:fidelity}
\end{equation}
This function is also
invariant under the interchange of Eq.~(\ref{eq:interchange})
and the rescaling of Eq.~(\ref{eq:rescale}).

The fidelity in Eq.~(\ref{eq:Fm}) is also at the level of a single outcome,
in the sense that it has its value
when a single outcome $m$ has been obtained.
If $F(m)$ is averaged over all outcomes
with probabilities given by Eq.~(\ref{eq:pm}),
the mean operation fidelity~\cite{Banasz01} is obtained:
\begin{equation}
   F\equiv\sum_m p(m)\,F(m)=\sum_m
     \overline{\left|\bra{\psi}\hat{M}_m\ket{\psi}\right|^2},
\label{eq:Fav}
\end{equation}
whose explicit form is given by~\cite{Banasz01}
\begin{equation}
    F =\frac{1}{d(d+1)}\left(d
                +\sum_m \tau_m^2 \right)
\end{equation}
from Eqs.~(\ref{eq:pm}), (\ref{eq:qmbar}),
and (\ref{eq:identity4normHS}),
though the average fidelity of Eq.~(\ref{eq:Fav})
is not discussed further in this paper.

\subsection{Physical reversibility}
Next, the degree of reversibility of the measurement is considered.
A quantum measurement is said to
be physically reversible~\cite{UeImNa96,Ueda97}
if the premeasurement state can be recovered from
the postmeasurement state with a nonzero probability of success
via a reversing measurement.
The necessary and sufficient condition for physical reversibility
is that the measurement operator $\hat{M}_m$
has a bounded left inverse $\hat{M}_m^{-1}$.
If this condition is satisfied,
then the reversing measurement can be constructed by
another set of measurement operators $\{\hat{R}_\mu^{(m)}\}$ that satisfy
\begin{equation}
\sum_\mu\hat{R}^{(m)\dagger}_\mu\hat{R}^{(m)}_\mu=\hat{I}
\label{eq:completenessRev}
\end{equation}
and in addition, for a particular $\mu=\mu_0$,
\begin{equation}
 \hat{R}^{(m)}_{\mu_0}=\kappa_m \hat{M}_m^{-1},
\end{equation}
where $\mu$ denotes the outcome of the reversing measurement
and $\kappa_m$ is a complex constant.
When the reversing measurement $\{\hat{R}_\mu^{(m)}\}$ is performed
on the postmeasurement state given in Eq.~(\ref{eq:reduction})
and the preferred outcome $\mu_0$ is obtained,
the state of the system successfully
reverts to the premeasurement state $\ket{\psi}$,
except for an overall phase factor via the second state reduction,
\begin{equation}
  \ket{\psi_{m\mu_0}}=
  \frac{1}{\sqrt{p_{m\mu_0}}}\,\hat{R}_{\mu_0}^{(m)}\ket{\psi_m}
  \propto \ket{\psi},
\end{equation}
where
\begin{equation}
  p_{m\mu_0} = \bra{\psi_m}\hat{R}_{\mu_0}^{(m)^\dagger}\hat{R}_{\mu_0}^{(m)}
            \ket{\psi_m}=\frac{|\kappa_m|^2}{p_m}
\end{equation}
is the probability for the second outcome $\mu_0$
given the first outcome $m$ and thus
is the successful probability of the reversing measurement.
Then, the physical reversibility can be evaluated using
the maximum successful probability
of the reversing measurement~\cite{KoaUed99,Ban01,KorJor06,CheLee12}.
Since the completeness condition
given in Eq.~(\ref{eq:completenessRev}) requires
$\bra{\psi}\hat{R}^{(m)\dagger}_{\mu_0}\hat{R}^{(m)}_{\mu_0}\ket{\psi}\le 1$
for any $\ket{\psi}$,
the upper bound for $|\kappa_m|^2$ is given by~\cite{KoaUed99}
\begin{equation}
|\kappa_m|^2\le
\inf_{\ket{\psi}}\,\bra{\psi} \hat{M}_m^\dagger\hat{M}_m \ket{\psi}
=\lambda_{m,\min}^2,
\end{equation}
where $\lambda_{m,\min}$ is the minimum singular value of $\hat{M}_m$,
\begin{equation}
       \lambda_{m,\min} \equiv \min_{j} \lambda_{mj}.
\label{eq:minimumValue}
\end{equation}
Therefore, the maximum successful probability
of the reversing measurement is
\begin{equation}
 \max_{\kappa_m}\, p_{m\mu_0}
    =\frac{\lambda_{m,\min}^2}{p_m},
\label{eq:maximumProbability}
\end{equation}
which is regarded as a measure of physical reversibility of measurement.

In this situation, 
when the measurement on the premeasurement state $\ket{\psi(a)}$
yields an outcome $m$,
the reversibility of Eq.~(\ref{eq:maximumProbability}) is given by
\begin{equation}
  R(m,a)
   =\frac{\lambda_{m,\min}^2}{p(m|a)}=\frac{\lambda_{m,\min}^2}{q_m(a)}
\end{equation}
on the basis of Eq.~(\ref{eq:pma}).
By averaging over the premeasurement states $\{\ket{\psi(a)}\}$
with probabilities given by Eq.~(\ref{eq:pam}),
the reversibility of the measurement with the outcome $m$ can be expressed as
\begin{equation}
  R(m) = \sum_a p(a|m)\,R(m,a)
       =d\left(\frac{\lambda_{m,\min}^2 }{\sigma_m^2}\right)
\label{eq:reversibility}
\end{equation}
by using Eq.~(\ref{eq:qmbar}).
This function is also invariant
under the interchange of Eq.~(\ref{eq:interchange})
and the rescaling of Eq.~(\ref{eq:rescale}).
Again, this reversibility is at the level of a single outcome
in the sense that it has its value when a single outcome $m$ has been obtained.
If $R(m)$ is averaged over all outcomes
with probabilities given by Eq.~(\ref{eq:pm}),
the degree of physical reversibility of a measurement
that was discussed in Ref.~\cite{KoaUed99} is obtained:
\begin{equation}
   R\equiv\sum_m p(m)\,R(m)
    =\sum_m \inf_{\ket{\psi}}\, \bra{\psi}\hcs{\hat{M}_m}\ket{\psi},
\label{eq:Rav}
\end{equation}
whose explicit form is given by~\cite{CheLee12}
\begin{equation}
    R =\sum_m \lambda_{m,\min}^2
\end{equation}
from Eqs.~(\ref{eq:pm}) and (\ref{eq:qmbar}),
though the average reversibility of Eq.~(\ref{eq:Rav})
is not discussed further in this paper.

\subsection{State estimation}
Finally, another measure of information gain
called estimation fidelity is introduced
to show its general formula at the level of a single outcome,
though this paper will mainly use $I(m)$ given in Eq.~(\ref{eq:Im}).
Suppose that when the measurement yields an outcome $m$,
the premeasurement state is estimated by a state $\ket{\varphi(m)}$.
If the actual premeasurement state is $\ket{\psi(a)}$,
the quality of the estimation can be evaluated
by the overlap $\bigl|\expv{\varphi(m)|\psi(a)}\bigr|^2$.
By averaging over the premeasurement states $\{\ket{\psi(a)}\}$
with probabilities given by Eq.~(\ref{eq:pam}),
the estimation fidelity after the measurement with the outcome $m$
can be expressed as
\begin{equation}
 G(m) =\sum_a p(a|m)\,\bigl|\expv{\varphi(m)|\psi(a)}\bigr|^2,
\end{equation}
which depends on the strategy of selecting $\ket{\varphi(m)}$.
In the optimal case~\cite{Banasz01},
the estimation $\ket{\varphi(m)}$ is assigned to
the eigenvector of $\hcs{\hat{M}_m}$ corresponding to
its maximum eigenvalue.
Since $\hcs{\hat{M}_m}=\hat{D}_m^2$ from Eq.~(\ref{eq:operator}),
$\ket{\varphi(m)}$ is one of the states in the basis $\{\ket{i}\}$;
namely, $\ket{\varphi(m)}=\ket{l}$,
with $l$ being one of $1,2,\ldots,d$ that satisfies
\begin{equation}
       \lambda_{ml} = \max_{j} \lambda_{mj}\equiv \lambda_{m,\max}.
\end{equation}
Using this strategy,
the estimation fidelity can be written as
\begin{equation}
 G(m) =\frac{1}{\;\overline{q_m}\;}
     \sum_{i} \lambda_{mi}^2
     \frac{1}{N}\sum_a \left| c_i(a)\right|^2\left| c_l(a)\right|^2,
\end{equation}
which is explicitly calculated to be
\begin{equation}
 G(m) =\frac{1}{d+1}\left(\frac{\sigma_m^2
                +\lambda_{m,\max}^2}{\sigma_m^2}\right)
\label{eq:estimationFidelity}
\end{equation}
using the calculations in Appendix \ref{sec:average}.
This function is also invariant
under the interchange of Eq.~(\ref{eq:interchange})
and the rescaling of Eq.~(\ref{eq:rescale}).

This estimation fidelity is at the level of a single outcome.
If $G(m)$ is averaged over all outcomes
with probabilities given by Eq.~(\ref{eq:pm}),
the mean estimation fidelity~\cite{Banasz01} is obtained:
\begin{equation}
  G\equiv\sum_m p(m)\,G(m)=\sum_m
     \overline{\bra{\psi}\hcs{\hat{M}_m}\ket{\psi}
\bigl|\expv{\varphi(m)|\psi}\bigr|^2},
\end{equation}
whose explicit form is given by~\cite{Banasz01}
\begin{equation}
  G =\frac{1}{d(d+1)}\left(d
                +\sum_m \lambda_{m,\max}^2 \right)
\end{equation}
from Eqs.~(\ref{eq:pm}), (\ref{eq:qmbar}),
and (\ref{eq:identity4normHS}).

\section{\label{sec:degeneracy}Degeneracy}
When some singular values are degenerate,
Eq.~(\ref{eq:information}) for information gain
is not useful for numerical calculations
due to the apparent divergences of
\begin{equation}
J\equiv \sum_{i}
  \frac{\lambda_{mi}^{2d}\log_2\lambda_{mi}^2}
    {\prod_{k\neq i}\left( \lambda_{mi}^2-\lambda_{mk}^2\right)}.
\label{eq:dangerous}
\end{equation}
Of course, $J$ is finite, because
it arises from the integral of a bounded function over a bounded region
as in Eq.~(\ref{eq:qmlogqmbar}).
Even if $\lambda_{mi}=\lambda_{mk}$,
a finite result can be obtained
by taking the limit as $\lambda_{mi}\to\lambda_{mk}$.
However, this limit operation is quite complicated
if singular values are highly degenerate.
Therefore, another formula will be presented for the information gain
that requires no limit operations
even when singular values are degenerate.

Since the ordering of singular values is insignificant
due to the invariance under the interchange of Eq.~(\ref{eq:interchange}),
they can first be divided into groups on the basis of their values:
\begin{equation}
  \left\{\lambda_{mi}\right\}
   \,\longrightarrow\,
  \left\{\left(\bar{\lambda}_{ms},n_s\right) \right\},
\label{eq:grouping}
\end{equation}
where the $s$th group contains $n_s$ singular values of $\bar{\lambda}_{ms}$,
and thus $\sum_s n_s=d$.
For example,
if the singular values are
\begin{equation}
\lambda_{m1}=\lambda_{m2}=\frac{1}{4}, \quad
\lambda_{m3}=\lambda_{m4}=\lambda_{m5}=\frac{1}{2}, \quad
\lambda_{m6}=\frac{3}{4},
\end{equation}
they are divided into three groups as
\begin{equation}
\left(\bar{\lambda}_{m1},n_1\right)=\left(\frac{1}{4},2\right), \quad
\left(\bar{\lambda}_{m2},n_2\right)=\left(\frac{1}{2},3\right), \quad
\left(\bar{\lambda}_{m3},n_3\right)=\left(\frac{3}{4},1\right).
\end{equation}
In accordance with this grouping,
the summation over $i$ in Eq.~(\ref{eq:dangerous})
can be expressed as a summation over the groups
\begin{equation}
  J=\sum_s J_s,
\label{eq:groupSum}
\end{equation}
where $J_s$ is the sum within
the $s$th group $\left(\bar{\lambda}_{ms},n_s\right)$
defined as a limit of
$\lambda_{1},\lambda_{2},\ldots,\lambda_{n_s}\to\bar{\lambda}_{ms}$:
\begin{equation}
  J_s=\lim_{\stackrel{\lambda_{1},\lambda_{2},\ldots,
       \lambda_{n_s}}{\to\bar{\lambda}_{ms}}}
    \sum_{i=1}^{n_s}\left(\prod_{ k\neq i}^{n_s}
      \frac{1}{\lambda_{i}^2-\lambda_{k}^2}\right)
       \frac{\lambda_{i}^{2d}\log_2\lambda_{i}^2}
  {\prod_{r\neq s}\left( \lambda_{i}^2
        -\bar{\lambda}_{mr}^2\right)^{n_{r}}}.
\end{equation}
This limit can be calculated as follows:
First, substitute $\bar{\lambda}_{ms}^2$ for $\lambda_{1}^2$
and $\bar{\lambda}_{ms}^2+\epsilon$ for $\lambda_{2}^2$,
and then take the limit as $\epsilon\to0$.
Next, substitute $\bar{\lambda}_{ms}^2+\epsilon$ for $\lambda_{3}^2$
and take the limit as $\epsilon\to0$.
Repeat similarly one by one
for $\lambda_{4}^2,\lambda_{5}^2,\ldots,\lambda_{n_s}^2$.
As a consequence of these procedures,
one find that at the last step $J_s$ should be of the form
\begin{equation}
  J_s=\lim_{\epsilon\to0}\left[
      \frac{1}{\epsilon^{n_s-1}}
      \frac{\left(\bar{\lambda}_{ms}^2+\epsilon\right)^d \log_2 
  \left(\bar{\lambda}_{ms}^2+\epsilon\right)}
  {\prod_{r\neq s}\left(\bar{\lambda}_{ms}^2+\epsilon
        -\bar{\lambda}_{mr}^2\right)^{n_{r}}}
   +\sum_{n=1}^{n_s-1}\frac{w_n^{(s)}}{\epsilon^n}\right],
\label{eq:lastLimitJs}
\end{equation}
where $\{w_n^{(s)}\}$ are finite coefficients.
Therefore, using the coefficients $\{z_n^{(s)}\}$ defined by Taylor series
\begin{equation}
       \frac{\left(\bar{\lambda}_{ms}^2+\epsilon\right)^d \log_2 
  \left(\bar{\lambda}_{ms}^2+\epsilon\right)}
  {\prod_{r\neq s}\left(\bar{\lambda}_{ms}^2+\epsilon
        -\bar{\lambda}_{mr}^2\right)^{n_{r}}}
   \equiv\sum_{n=0}^\infty z_n^{(s)}\,\epsilon^n,
\label{eq:zsnDef}
\end{equation}
$J_s$ can be written with no limit operations as
\begin{equation}
  J_s=z_{n_s-1}^{(s)}.
\label{eq:Js2zsn}
\end{equation}
Note that when Eq.~(\ref{eq:zsnDef}) is substituted
into Eq.~(\ref{eq:lastLimitJs}),
the divergent terms containing $\{w_n^{(s)}\}$ with $n=1,2,\ldots,n_s-1$
should be canceled by the divergent terms containing
$\{z_n^{(s)}\}$ with $n=n_s-2,n_s-3,\ldots,0$,
since $J_s$ is finite.

A more explicit form of $J_s$ can be found by
separating the left-hand side of Eq.~(\ref{eq:zsnDef}) into two parts
that can then be expanded as Taylor series.
The first part is
\begin{equation}
\left(\lambda^2+\epsilon\right)^d \log_2 
  \left(\lambda^2+\epsilon\right) \equiv
\sum_{n=0}^{d-1} c^{(d)}_n(\lambda)\, \epsilon^n
 +O\left(\epsilon^d\right),
\label{eq:cdnDef}
\end{equation}
which corresponds to the numerator of Eq.~(\ref{eq:zsnDef}).
As shown in Appendix \ref{sec:coefficient},
the coefficients $\{c^{(d)}_n(\lambda)\}$ for $n=0,1,\ldots,d-1$
are given by
\begin{equation}
c^{(d)}_n(\lambda) = \lambda^{2(d-n)}
   \left[\binom{d}{n}\log_2 \lambda^2+a^{(d)}_n\right],
\label{eq:cdn}
\end{equation}
where the coefficients $\{a^{(d)}_n\}$ are
\begin{equation}
a^{(d)}_n =\frac{1}{\ln2}\binom{d}{n}
       \Bigl[\eta(d)- \eta(d-n)\Bigr].
\label{eq:adn}
\end{equation}
The explicit forms of $\{a^{(d)}_n\}$ are
\begin{align}
 & a^{(d)}_0 = 0, \quad a^{(d)}_1 = \frac{1}{\ln2},\quad
   a^{(d)}_2 = \frac{1}{\ln2}\left(d-\frac{1}{2}\right), \notag \\
 & \qquad\qquad\ldots,\quad
 a^{(d)}_{d-1} = \frac{d}{\ln2}\left(\frac{1}{2}+\cdots+\frac{1}{d}\right).
\end{align}
It is clear that
\begin{equation}
  c^{(d)}_n(0)=0, \qquad c^{(d)}_n(1)=a^{(d)}_n.
\end{equation}
On the other hand, the second part is
\begin{equation}
 \frac{1}{\prod_{r\neq s}\left(\bar{\lambda}_{ms}^2+\epsilon
        -\bar{\lambda}_{mr}^2\right)^{n_{r}}}
   \equiv\frac{1}{\prod_{r\neq s}\left(\bar{\lambda}_{ms}^2
        -\bar{\lambda}_{mr}^2\right)^{n_{r}}}\,
  \sum_{n=0}^{\infty}\, b_{n}^{(s)}\,\epsilon^n.
\label{eq:bsnDef}
\end{equation}
The coefficients $\{b_{n}^{(s)}\}$ are complicated in general,
but they can be described in a compact form
with the help of complete Bell polynomials
\begin{equation}
B_{n}\left(x_1,x_2,\ldots,x_{n}\right)= 
\sum_{\{j_r\}} \frac{n!}{j_1!j_2!\cdots j_{n}!}
   \left(\frac{x_1}{1!}\right)^{j_1}\left(\frac{x_2}{2!}\right)^{j_2}
   \cdots\left(\frac{x_{n}}{n!}\right)^{j_{n}},
\label{eq:BellPolynomial}
\end{equation}
where the summation is taken
over all possible sets of non-negative integers $\{j_r\}$ such that
\begin{equation}
  \sum_{r=1}^n r j_r=n.
\end{equation}
The explicit forms for $n=0$, $1$, $2$, and $3$ are
\begin{align}
 & B_0 = 1, \notag \\
 & B_1(x_1) = x_1, \notag \\
 & B_2(x_1,x_2) = x_1^2+x_2, \notag \\
 & B_3(x_1,x_2,x_3) = x_1^3+3x_1x_2+x_3.
\end{align}
With these complete Bell polynomials,
the coefficients $\{b_{n}^{(s)}\}$ are given by
\begin{equation}
  b_{n}^{(s)}=\frac{1}{n!}\, B_{n}\left(h^{(s)}_1,h^{(s)}_2,
           \ldots,h^{(s)}_{n}\right),
\label{eq:bsn}
\end{equation}
where the coefficients $\{h^{(s)}_n\}$ are
\begin{equation}
h^{(s)}_n =(-1)^n(n-1)! \sum_{r\neq s}
   \frac{n_{r}}{\left(\bar{\lambda}_{ms}^2
        -\bar{\lambda}_{mr}^2\right)^{n}},
\label{eq:hsn}
\end{equation}
as shown in Appendix \ref{sec:coefficient}.
By substituting
the Taylor series of Eqs.~(\ref{eq:cdnDef}) and (\ref{eq:bsnDef})
into Eq.~(\ref{eq:zsnDef}),
Eq.~(\ref{eq:Js2zsn}) can be expressed as
\begin{equation}
  J_s=\frac{1}{\prod_{r\neq s}\left(\bar{\lambda}_{ms}^2
        -\bar{\lambda}_{mr}^2\right)^{n_{r}}}
   \sum_{n=0}^{n_s-1}
   c^{(d)}_{n}(\bar{\lambda}_{ms})\, b_{n_s-1-n}^{(s)}.
\label{eq:Js}
\end{equation}
Performing the summation of Eq.~(\ref{eq:groupSum}) over all groups
then yields
\begin{equation}
  J=\sum_s \frac{1}{\prod_{r\neq s}\left(\bar{\lambda}_{ms}^2
        -\bar{\lambda}_{mr}^2\right)^{n_{r}}}
   \sum_{n=0}^{n_s-1}
c^{(d)}_{n}(\bar{\lambda}_{ms})\, b_{n_s-1-n}^{(s)}.
\label{eq:J}
\end{equation}
Since $\bar{\lambda}_{ms} \neq \bar{\lambda}_{mr}$ if $s\neq r$
due to the grouping of Eq.~(\ref{eq:grouping}),
this expression is clearly free from apparent divergences,
thus eliminating the need for limit operations
even when the singular values are degenerate.
In particular,
Eq.~(\ref{eq:J}) is more useful than Eq.~(\ref{eq:dangerous})
for numerical calculations,
by which the author has verified the consistency of
Eq.~(\ref{eq:information}) with Eq.~(\ref{eq:Im})
by using the Monte Carlo method for integration.

To outline the calculation of Eq.~(\ref{eq:J}),
a simple case is presented wherein
the singular values in $d=6$ are divided into three groups:
\begin{equation}
\left(\bar{\lambda}_{m1},n_1\right)=\left(\lambda,3\right), \quad
\left(\bar{\lambda}_{m2},n_2\right)=\left(\sqrt{2}\lambda,2\right), \quad
\left(\bar{\lambda}_{m3},n_3\right)=\left(\sqrt{3}\lambda,1\right).
\end{equation}
The first group $s=1$ can be used to obtain $J_1$.
Since $n_1=3$, it is necessary to calculate $b^{(1)}_0$, $b^{(1)}_1$,
and $b^{(1)}_2$  from Eq.~(\ref{eq:Js}),
which themselves require $h^{(1)}_1$ and $h^{(1)}_2$, as in Eq.~(\ref{eq:bsn}).
According to Eq.~(\ref{eq:hsn}),
\begin{equation}
h^{(1)}_1 =\frac{5}{2\lambda^2}, \qquad
h^{(1)}_2 =\frac{9}{4\lambda^4},
\end{equation}
which gives
\begin{equation}
b^{(1)}_0 =1, \qquad
b^{(1)}_1 =\frac{5}{2\lambda^2}, \qquad
b^{(1)}_2 =\frac{17}{4\lambda^4}.
\end{equation}
By combining these coefficients with
$c^{(6)}_{0}(\lambda)$, $c^{(6)}_{1}(\lambda)$,
and $c^{(6)}_{2}(\lambda)$,
the following equation can be obtained:
\begin{equation}
J_1 = -\lambda^2 \left(\frac{137}{8}\log_2 \lambda^2+
         \frac{4}{\ln 2}\right).
\end{equation}
Similar calculations should be done
for the second and third groups, $s=2$ and $s=3$,
to obtain $J_2$ and $J_3$.
Then, $J$ can be obtained by adding $J_1$, $J_2$, and $J_3$,
though the result is omitted here.

\section{\label{sec:example}Example}
As an example, a class of quantum measurements
with highly degenerate singular values is considered
next to discuss tradeoffs among
the information, fidelity, and reversibility
that are given by Eqs.~(\ref{eq:information}), (\ref{eq:fidelity}),
and (\ref{eq:reversibility}), respectively.
The measurement considered here is described
by a measurement operator whose singular values are
\begin{align}
 & \lambda_{m1} =\lambda_{m2} =\cdots=\lambda_{mk}=1,  \notag \\
& \lambda_{m(k+1)} =\lambda_{m(k+2)} =\cdots=\lambda_{m(k+l)}=\lambda,
\label{eq:measurementEx} \\
& \lambda_{m(k+l+1)} =\lambda_{m(k+l+2)} =\cdots=\lambda_{md}=0, \notag
\end{align}
when it yields an outcome $m$.
The singular values are sorted in descending order
by the interchange of Eq.~(\ref{eq:interchange}),
and the maximum singular values are normalized to $1$
by the rescaling of Eq.~(\ref{eq:rescale}).
Note that if $k=0$, $l=0$, $\lambda=0$, or $\lambda=1$,
this measurement becomes a projective measurement,
as was discussed in Ref.~\cite{Terash11b}.
Therefore, it is assumed that
\begin{equation}
 k=1,2,\ldots, d-1, \quad l=1,2,\ldots,d-k,
  \quad 0 < \lambda < 1.
\end{equation}

First, the calculation of the information
given by Eq.~(\ref{eq:information}) is presented
with dividing the singular values into groups
as in Eq.~(\ref{eq:grouping}) to handle their degeneracies:
\begin{equation}
\left(\bar{\lambda}_{m1},n_1\right)=\left(1,k\right), \quad
\left(\bar{\lambda}_{m2},n_2\right)=\left(\lambda,l\right), \quad
\left(\bar{\lambda}_{m3},n_3\right)=\left(0,d-k-l\right).
\end{equation}
In this case,
Eq.~(\ref{eq:Js2zsn}) should be used rather than Eq.~(\ref{eq:Js})
to calculate the dangerous term of Eq.~(\ref{eq:dangerous}),
because it is easy
to expand the left-hand side of Eq.~(\ref{eq:zsnDef})
as a Taylor series.
In fact, for the first group $s=1$,
Eq.~(\ref{eq:zsnDef}) becomes
\begin{equation}
\frac{\left(1+\epsilon\right)^{k+l} \log_2 
  \left(1+\epsilon\right)}
  {\left(1+\epsilon-\lambda^2\right)^{l}}
  =\sum_{n=0}^{\infty} z_n^{(1)}\,\epsilon^n.
\end{equation}
The numerator is expanded as in Eq.~(\ref{eq:cdnDef}),
with coefficients $c^{(k+l)}_n(1)=a^{(k+l)}_n$,
while the remaining part can be expanded
by the generalized binomial theorem as
\begin{equation}
\frac{1}
  {\left(1+\epsilon-\lambda^2\right)^{l}}
 = (-1)^l\sum_{n=0}^{\infty}
    \binom{l+n-1}{l-1}\frac{1}
  {\left(\lambda^2-1\right)^{l+n}} \,\epsilon^n.
\end{equation}
Using these Taylor series, for the first group $s=1$,
$J_1$ can be found to be
\begin{equation}
    J_1=(-1)^l \sum_{n=0}^{k-1}\binom{k+l-n-2}{l-1}\,
     \frac{a^{(k+l)}_n}{(\lambda^2-1)^{k+l-n-1}}
\label{eq:dangerousJ1}
\end{equation}
and, similarly, for the second group $s=2$,
\begin{equation}
    J_2=(-1)^k\sum_{n=0}^{l-1}\binom{k+l-n-2}{k-1}\,
     \frac{c^{(k+l)}_n(\lambda)}{(1-\lambda^2)^{k+l-n-1}}.
\label{eq:dangerousJ2}
\end{equation}
On the other hand, for the third group $s=3$,
$J_3=0$ can be obtained from Eq.~(\ref{eq:Js}) because $c^{(d)}_n(0)=0$.
The dangerous term in Eq.~(\ref{eq:dangerous}) is then
given by $J=J_1+J_2$, as in Eq.~(\ref{eq:groupSum}).
From the resultant $J$,
the information of Eq.~(\ref{eq:information}) can be calculated as
\begin{align}
 &I(m) = \log_2d
  -\frac{1}{\ln2}\Bigl[\eta(d)- 1\Bigr]-
     \log_2\left(k+l\lambda^2\right)
\notag\\
&\qquad
+\frac{1}{k+l\lambda^2}\left[
(-1)^l \sum_{n=0}^{k-1}\binom{k+l-n-2}{l-1}\,
     \frac{a^{(k+l)}_n}{(\lambda^2-1)^{k+l-n-1}}
\right.
\notag\\
&\qquad\qquad\qquad 
\left.
+(-1)^k\sum_{n=0}^{l-1}\binom{k+l-n-2}{k-1}\,
     \frac{c^{(k+l)}_n(\lambda)}{(1-\lambda^2)^{k+l-n-1}}
\right],
\label{eq:informationEx}
\end{align}
since the Hilbert--Schmidt norm of Eq.~(\ref{eq:normHS})
is $\sigma_m^2=k+l\lambda^2$ in this case.
\begin{figure}
\begin{center}
\includegraphics[scale=0.6]{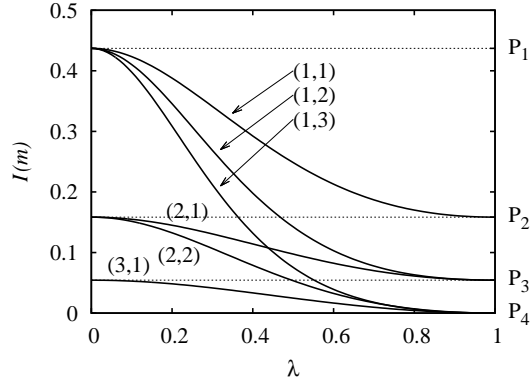}
\end{center}
\caption{\label{fig1}
Information $I(m)$
as a function of singular value $\lambda$ in $d=4$
for $(k,l)=(1,1)$, $(1,2)$, $(1,3)$, $(2,1)$, $(2,2)$, and $(3,1)$.
The symbols $\{\mathrm{P}_{r}\}$ $(r=1,2,3,4)$
denote projective measurements of rank $r$.}
\end{figure}%
Figure~\ref{fig1} shows this information $I(m)$
as a function of $\lambda$ in $d=4$ for various $(k,l)$.
In the figure, the symbols $\{\mathrm{P}_{r}\}$ $(r=1,2,3,4)$
denote projective measurements of rank $r$,
even though $\mathrm{P}_{4}$ in $d=4$
is nothing more than the identity operation.
The information for $\mathrm{P}_{r}$ is given by~\cite{Terash11b}
\begin{equation}
  I(m)=\log_2\frac{d}{r}-\frac{1}{\ln2}\Bigl[\eta(d)- \eta(r)\Bigr].
\label{eq:informationProj}
\end{equation}
As shown in Fig.~\ref{fig1},
the information of Eq.~(\ref{eq:informationEx}) for $(k,l)$
is equal to that for $\mathrm{P}_{k}$ when $\lambda=0$
and is equal to that for $\mathrm{P}_{k+l}$ when $\lambda=1$,
as expected;
these facts can be confirmed mathematically
from Eq.~(\ref{eq:informationEx})
as shown in Appendix \ref{sec:limits}.
The estimation fidelity $G(m)$ given in Eq.~(\ref{eq:estimationFidelity})
also changes in a similar way to $I(m)$ between $1/4$ and $2/5$.

At the same time,
the fidelity of Eq.~(\ref{eq:fidelity}) and
reversibility of Eq.~(\ref{eq:reversibility}) can be
calculated to be
\begin{equation}
 F(m) =
    \frac{1}{d+1}
   \left[\frac{k(k+1)+2kl\lambda+l(l+1)\lambda^2}{k+l\lambda^2}\right]
\label{eq:fidelityEx}
\end{equation}
and 
\begin{equation}
  R(m) =
     d\left(\frac{\lambda^2}{k+l\lambda^2}\right)\delta_{d,(k+l)},
\label{eq:reversibilityEx}
\end{equation}
respectively,
since the trace norm of Eq.~(\ref{eq:normT}) is $\tau_m=k+l\lambda$ and
the minimum singular value of Eq.~(\ref{eq:minimumValue}) is
$\lambda_{m,\min}=\lambda\delta_{d,(k+l)}$.
\begin{figure}
\begin{center}
\includegraphics[scale=0.6]{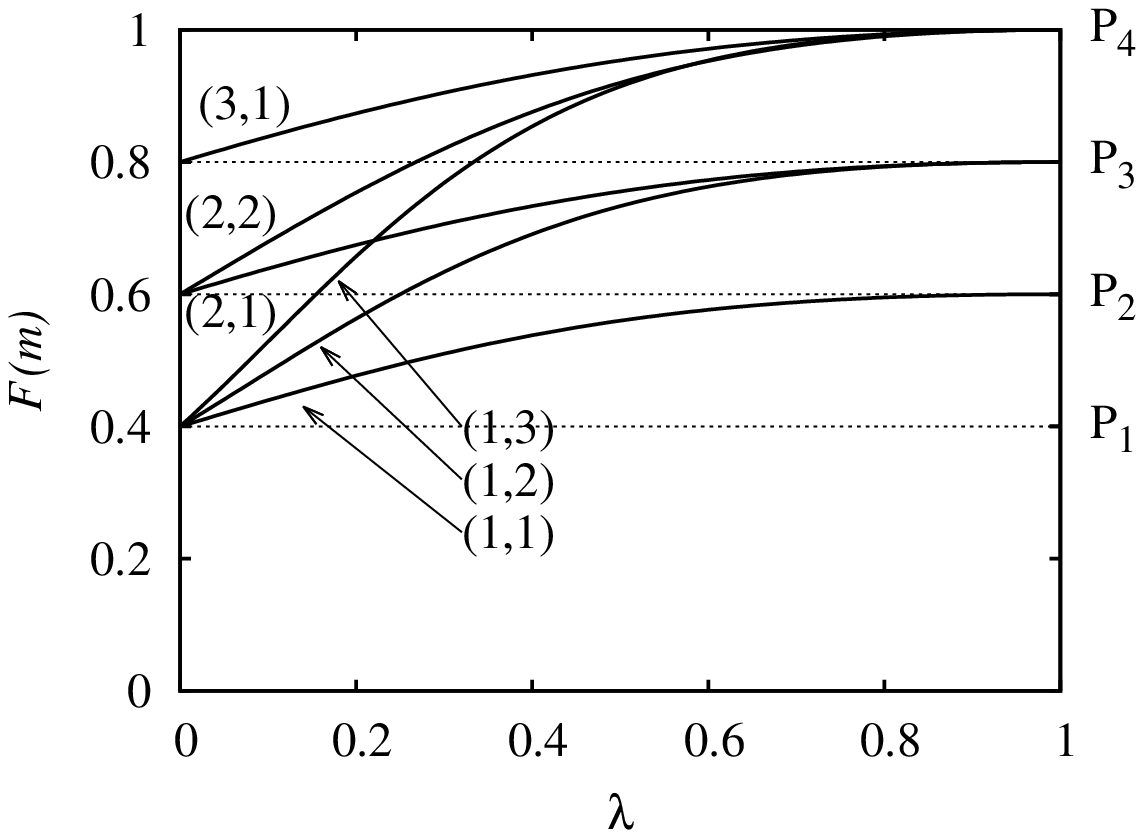}
\end{center}
\caption{\label{fig2}
Fidelity $F(m)$
as a function of singular value $\lambda$ in $d=4$
for $(k,l)=(1,1)$, $(1,2)$, $(1,3)$, $(2,1)$, $(2,2)$, and $(3,1)$.
The symbols $\{\mathrm{P}_{r}\}$ $(r=1,2,3,4)$
denote projective measurements of rank $r$.}
\end{figure}%
\begin{figure}
\begin{center}
\includegraphics[scale=0.6]{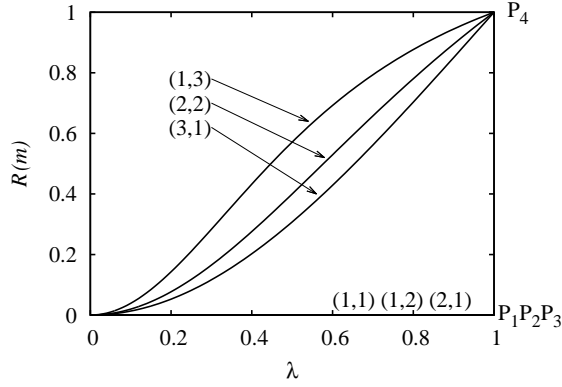}
\end{center}
\caption{\label{fig3}
Reversibility $R(m)$
as a function of singular value $\lambda$ in $d=4$
for $(k,l)=(1,1)$, $(1,2)$, $(1,3)$, $(2,1)$, $(2,2)$, and $(3,1)$.
The symbols $\{\mathrm{P}_{r}\}$ $(r=1,2,3,4)$
denote projective measurements of rank $r$.}
\end{figure}%
Figures~\ref{fig2} and \ref{fig3} show this fidelity $F(m)$
and reversibility $R(m)$
as functions of $\lambda$ in $d=4$ for various $(k,l)$,
while those for $\mathrm{P}_{r}$ are given by~\cite{Terash11b}
\begin{equation}
 F(m) = \frac{r+1}{d+1}
\label{eq:fidelityProj}
\end{equation}
and 
\begin{equation}
  R(m) =\delta_{d,r}.
\label{eq:reversibilityProj}
\end{equation}
The reversibility of Eq.~(\ref{eq:reversibilityEx})
is $0$ for each of $(k,l)=(1,1)$, $(1,2)$, and $(2,1)$
since $\lambda_{m,\min}=0$, as shown in Fig.~\ref{fig3}.

Now the tradeoffs among information gain,
state change, and physical reversibility can be discussed
for this class of measurements,
since the three quantities have been expressed as
functions of the same single parameter $\lambda$.
As the parameter $\lambda$ increases,
the information of Eq.~(\ref{eq:informationEx}) monotonically decreases,
as in Fig.~\ref{fig1},
whereas the fidelity of Eq.~(\ref{eq:fidelityEx}) and
reversibility of Eq.~(\ref{eq:reversibilityEx})
monotonically increase, as in Figs.~\ref{fig2} and \ref{fig3}.
Thus, as a measurement provides
more information about the state of the system,
it changes the state less reversibly and to a greater extent.
Therefore, loss of fidelity and loss of reversibility
are both regarded as costs of information gain.

To explore the balance between costs and gains,
two kinds of measurement efficiencies can be defined:
one is the ratio of information gain to fidelity loss,
\begin{equation}
    E_F(m)\equiv \frac{I(m)}{1-F(m)},
\label{eq:EFDef}
\end{equation}
and the other is
the ratio of information gain to reversibility loss,
\begin{equation}
    E_R(m)\equiv \frac{I(m)}{1-R(m)}.
\label{eq:ERDef}
\end{equation}
\begin{figure}
\begin{center}
\includegraphics[scale=0.6]{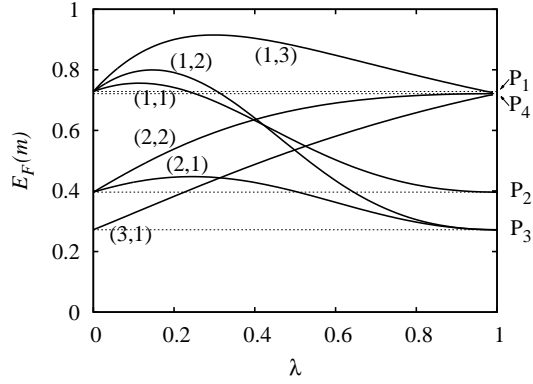}
\end{center}
\caption{\label{fig4}
Efficiency with respect to fidelity, $E_F(m)$,
as a function of singular value $\lambda$ in $d=4$
for $(k,l)=(1,1)$, $(1,2)$, $(1,3)$, $(2,1)$, $(2,2)$, and $(3,1)$.
The symbols $\{\mathrm{P}_{r}\}$ $(r=1,2,3,4)$
denote projective measurements of rank $r$.}
\end{figure}
\begin{figure}
\begin{center}
\includegraphics[scale=0.6]{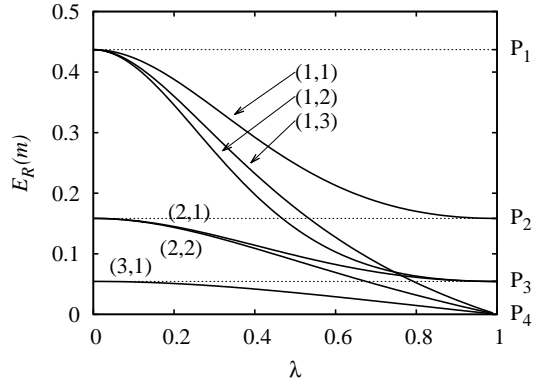}
\end{center}
\caption{\label{fig5}
Efficiency with respect to reversibility, $E_R(m)$,
as a function of singular value $\lambda$ in $d=4$
for $(k,l)=(1,1)$, $(1,2)$, $(1,3)$, $(2,1)$, $(2,2)$, and $(3,1)$.
The symbols $\{\mathrm{P}_{r}\}$ $(r=1,2,3,4)$
denote projective measurements of rank $r$.}
\end{figure}%
Figures~\ref{fig4} and \ref{fig5} show these efficiencies,
$E_F(m)$ and $E_R(m)$,
as functions of $\lambda$ in $d=4$ for various $(k,l)$.
As shown in Fig.~\ref{fig4},
the efficiency $E_F(m)$ is not always a monotonic function,
though it is difficult to analytically find its extreme value.
In contrast, as shown in Fig.~\ref{fig5},
the efficiency $E_R(m)$ is a monotonic function like
the information function $I(m)$.
In fact, for $(k,l)=(1,1)$, $(1,2)$, and $(2,1)$,
the efficiency $E_R(m)$ is identical to the information $I(m)$
because of the irreversibility, $R(m)=0$.
The efficiencies $E_F(m)$ and $E_R(m)$ for $\mathrm{P}_{r}$
can also be calculated from Eqs.~(\ref{eq:informationProj}),
(\ref{eq:fidelityProj}), and (\ref{eq:reversibilityProj})
when $r=1$, $2$, or $3$.
However, it is not straightforward to calculate the efficiencies
for the identity operation $\mathrm{P}_{4}$,
since $I(m)=0$ and $F(m)=R(m)=1$.
The limit values at $\mathrm{P}_{4}$ can be calculated
by considering the measurement of Eq.~(\ref{eq:measurementEx})
with $(k,l)=(d-1,1)$ and $\lambda^2=1-\epsilon$.
In this case, the information, fidelity, and reversibility
given by Eqs.~(\ref{eq:informationEx}), (\ref{eq:fidelityEx}),
and (\ref{eq:reversibilityEx}), respectively, can be expanded as
\begin{align}
  I(m) &= \frac{1}{2d^2\ln2}\left(\frac{d-1}{d+1}\right) \epsilon^2 
         +O\left(\epsilon^3\right),  \\
  F(m) &= 1-\frac{1}{4d}\left(\frac{d-1}{d+1}\right) \epsilon^2 
         +O\left(\epsilon^3\right), \\
  R(m) &= 1-\frac{d-1}{d} \epsilon -\frac{d-1}{d^2} \epsilon^2
         +O\left(\epsilon^3\right).
\end{align}
By taking the limit as $\epsilon\to 0$,
the limits of the efficiencies $E_F(m)$ and $E_R(m)$
at $\mathrm{P}_{4}$ are found to be
\begin{equation}
  E_F(m) \to \frac{2}{d\ln2},  \qquad
  E_R(m) \to 0.
\end{equation}

\section{\label{sec:conclude}Conclusion}
The information, fidelity, and reversibility
of an arbitrary quantum measurement
have been shown in a $d$-level system
whose premeasurement state is assumed to be completely unknown.
These quantities have been expressed as functions of
the singular values $\{\lambda_{mi}\}$
of the measurement operator $\hat{M}_m$
corresponding to the outcome $m$ of the measurement,
as shown in Eqs.~(\ref{eq:information}),
(\ref{eq:fidelity}), and (\ref{eq:reversibility}).
Unfortunately, when some singular values are degenerate,
Eq.~(\ref{eq:information}) for the information gain
is not useful due to the apparent divergence of
the dangerous term shown in Eq.~(\ref{eq:dangerous}).
Therefore, another expression for the dangerous term
was presented in Eq.~(\ref{eq:J}),
which is free of an apparent divergence
even when singular values are degenerate.
As an example, a class of quantum measurements was considered
whose singular values, as shown in Eq.~(\ref{eq:measurementEx}),
are highly degenerate.
According to the general formulas,
the information, fidelity, and reversibility
were calculated as shown in Eqs.~(\ref{eq:informationEx}),
(\ref{eq:fidelityEx}), and (\ref{eq:reversibilityEx}), respectively.
For $d=4$, these quantities are shown
in Figs.~\ref{fig1}, \ref{fig2}, and \ref{fig3},
which indicate the tradeoffs among
the information, fidelity, and reversibility.
That is,
as a measurement provides more information about the state of the system,
it changes the state by a greater degree and more irreversibly.
Two measurement efficiencies were also defined,
as shown in Eqs.~(\ref{eq:EFDef}) and (\ref{eq:ERDef}),
to show their different behaviors.

The formulas shown in this paper are applicable to
any efficient quantum measurement
in systems with a finite-dimensional Hilbert space,
such as multiple qubits or a qudit
in quantum information theory.
When an outcome is obtained by measurements,
it is possible to calculate how much information is provided
and how greatly and reversibly the state of the system is changed
directly from the singular values of the measurement operator
corresponding to the obtained outcome
with no optimization problems~\cite{Banasz01,BanDev01,Sacchi06}.
The three quantities are for each single outcome
rather than those averaged over all possible outcomes
with probabilities given by Eq.~(\ref{eq:pm}),
as shown in Eqs.~(\ref{eq:Iav}), (\ref{eq:Fav}), and (\ref{eq:Rav}).
It is not necessary to know the measurement operators 
corresponding to other outcomes.
Therefore, the tradeoffs
at the level of a single outcome are more fundamental
in quantum measurement.
Although the tradeoffs were shown only
in a specific class of measurements in this paper,
a general theory for such tradeoffs will be presented in future studies.
For general measurements, increasing information does not necessarily
result in decreasing fidelity or reversibility.
This is because the three quantities are functions of
$d-1$ parameters and hence
their relations are expressed by regions of finite size rather than lines.
However, the boundaries of the regions show tradeoffs
among information, fidelity, and reversibility.

\appendix
\section*{Appendix}

\section{\label{sec:average}Averages over States}
Herein, the averages of $q_m(a)$ 
and $\left[f_m(a)\right]^2$ over
all possible pure states of a $d$-level system are shown
to prove Eqs.~(\ref{eq:qmbar}) and (\ref{eq:fm2bar}).
They are given by
\begin{align}
\overline{q_m}&=\frac{1}{N}\sum_a
  \sum_{i} \lambda_{mi}^2\left| c_i(a)\right|^2, \label{eq:qmbarDef} \\
\overline{f_m^2}&=\frac{1}{N}\sum_a
  \sum_{i,j} \lambda_{mi}\lambda_{mj}
    \left| c_i(a)\right|^2\left| c_j(a)\right|^2, \label{eq:fm2barDef}
\end{align}
from Eqs.~(\ref{eq:pma}) and (\ref{eq:fma}),
together with Eq.~(\ref{eq:overline}).
First, the constants $C$, $D$, and $E$ can be defined as
\begin{equation}
\frac{1}{N}\sum_a \left| c_i(a)\right|^2\equiv C
\end{equation}
and
\begin{equation}
 \frac{1}{N}\sum_a
        |c_i(a)|^2|c_j(a)|^2 \equiv
   \begin{cases}
      D & \mbox{(if $i=j$)}; \\
      E & \mbox{(if $i\neq j$)}.
   \end{cases}
\end{equation}
Note that these constants do not depend on $i$ or $j$,
because there is no preferred state $\ket{i}$
when the index $a$ runs over all pure states of the system.
Using these constants,
Eqs.~(\ref{eq:qmbarDef}) and (\ref{eq:fm2barDef}) can be written as
\begin{align}
\overline{q_m}&=C \sum_{i} \lambda_{mi}^2=C\,\sigma_m^2,\label{eq:qmC} \\
\overline{f_m^2}&=D\sum_i\lambda_{mi}^2
            +E\sum_{i\neq j}\lambda_{mi}\lambda_{mj}
  =(D-E)\,\sigma_m^2+E\,\tau_m^2,   \label{eq:fm2DE}
\end{align}
where $\sigma_m$ and $\tau_m$ are defined by
Eqs.~(\ref{eq:normHS}) and (\ref{eq:normT}), respectively.

To calculate the constants $C$, $D$, and $E$,
a parameterization of the coefficients $\{c_i(a)\}$ can be introduced.
If $\alpha_i(a)$ and $\beta_i(a)$ are
the real and imaginary parts of $c_i(a)$, respectively,
then the normalization condition of Eq.~(\ref{eq:normalization}) becomes
\begin{equation}
   \sum_i \left[\alpha_i(a)^2+\beta_i(a)^2\right]=1,
\end{equation}
which is the condition for a point to be on
the unit sphere in $2d$ dimensions.
Thus, $\{\alpha_i(a)\}$ and $\{\beta_i(a)\}$ should be
parameterized by the hyperspherical coordinates
$(\theta_1,\theta_2,\ldots,\theta_{2d-2},\phi)$ as
\begin{align}
   \alpha_1(a) &= \sin\theta_{2d-2}\sin\theta_{2d-3}\cdots
                   \sin\theta_3\sin\theta_2\sin\theta_1\cos\phi, \notag \\
   \beta_1(a) &= \sin\theta_{2d-2}\sin\theta_{2d-3}\cdots
                   \sin\theta_3\sin\theta_2\sin\theta_1\sin\phi, \notag  \\
   \alpha_2(a) &= \sin\theta_{2d-2}\sin\theta_{2d-3}\cdots
                   \sin\theta_3\sin\theta_2\cos\theta_1, \notag  \\
   \beta_2(a) &= \sin\theta_{2d-2}\sin\theta_{2d-3}\cdots
                   \sin\theta_3\cos\theta_2,                    \\
             &\vdots  \notag \\
   \alpha_d(a) &= \sin\theta_{2d-2}\cos\theta_{2d-3}, \notag  \\
   \beta_d(a) &= \cos\theta_{2d-2},\notag 
\end{align}
where $0\le \phi < 2\pi$ and
$0\le \theta_p \le \pi$ for $p=1,2,\ldots,2d-2$.
The index $a$ can be replaced with
the angles $(\theta_1,\theta_2,\ldots,\theta_{2d-2},\phi)$,
and the summation over $a$ can be replaced with
the integral over the angles:
\begin{equation}
   \frac{1}{N}\sum_a   \quad\longrightarrow\quad
     \frac{(d-1)!}{2\pi^{d}}\int^{2\pi}_0 d\phi\,
     \prod^{2d-2}_{p=1}\int^\pi_0 d\theta_p\sin^p \theta_p.
\end{equation}
Then, if $i=1$ and $j=d$,
\begin{align}
 C &=\frac{1}{N}\sum_a \left| c_1(a)\right|^2
  =\frac{(d-1)!}{\pi^{d-1}}\,
   \prod^{2d-2}_{p=1}\int^\pi_0 d\theta_p\sin^{p+2} \theta_p, \\
 D &=\frac{1}{N}\sum_a \left| c_1(a)\right|^4 
   = \frac{(d-1)!}{\pi^{d-1}}\,
     \prod^{2d-2}_{p=1}\int^\pi_0 d\theta_p\sin^{p+4} \theta_p, \\
 E &=\frac{1}{N}\sum_a \left| c_1(a)\right|^2\left| c_d(a)\right|^2 \notag \\
&=  C-\frac{(d-1)!}{\pi^{d-1}}\,
     \prod^{2d-2}_{p=2d-3}\int^\pi_0 d\theta_p\sin^{p+4} \theta_p
     \times\prod^{2d-4}_{p=1}\int^\pi_0 d\theta_p\sin^{p+2} \theta_p.
\end{align}
These integrals can easily be calculated to be
\begin{equation}
  C = \frac{1}{d},  \qquad
  D = \frac{2}{d(d+1)},  \qquad
  E = \frac{1}{d(d+1)}
\label{eq:CDE}
\end{equation}
by using the integral formula
\begin{equation}
  \int^\pi_0 d\theta \,\sin^n \theta =\sqrt{\pi}\,
  \frac{\Gamma\left(\frac{n+1}{2}\right)}
      {\Gamma\left(\frac{n+2}{2}\right)}
\end{equation}
for $n>-1$ with the Gamma function $\Gamma(n)$.
Therefore,
Eqs.~(\ref{eq:qmbar}) and (\ref{eq:fm2bar}) can be proven
by substituting Eq.~(\ref{eq:CDE}) into
Eqs.~(\ref{eq:qmC}) and (\ref{eq:fm2DE}).

\section{\label{sec:coefficient}Coefficients of Series}
Herein, the coefficients of the Taylor series
in Eqs.~(\ref{eq:cdnDef}) and (\ref{eq:bsnDef}) are presented.
To find the coefficients $\{c^{(d)}_n(\lambda)\}$ in Eq.~(\ref{eq:cdnDef}),
the following Taylor series is first considered:
\begin{equation}
\left(1+\epsilon\right)^d \log_2 
  \left(1+\epsilon\right) \equiv
\sum_{n=0}^{d-1} a^{(d)}_n\,\epsilon^n
 +O\left(\epsilon^d\right).
\label{eq:adnDef}
\end{equation}
By expanding $\left(1+\epsilon\right)^d$ and
$\log_2 \left(1+\epsilon\right)$ in the Taylor series,
the coefficients $\{a^{(d)}_n\}$ can be determined to be
$a^{(d)}_0 =0$ for $n=0$ and 
\begin{equation}
   a^{(d)}_n =\frac{1}{\ln2}\sum_{k=1}^{n}
     \frac{(-1)^{k+1}}{k}\binom{d}{n-k}
\label{eq:adn2}
\end{equation}
for $n=1,2,\ldots,d-1$.

Next, a proof of the equivalence
between Eqs.~(\ref{eq:adn2}) and (\ref{eq:adn}) will be presented
by mathematical induction.
As the first step, it will be shown that
the statement holds for $a^{(d)}_{1}$ and $a^{(d)}_{d-1}$.
It is easy to see that both equations yield $a^{(d)}_{1}=1/\ln 2$.
At the same time, using the identity
\begin{equation}
  \frac{1}{k}\binom{d}{d-1-k}
  =d\left[\frac{1}{k}-\frac{1}{k+1}\right]\binom{d-1}{k}
\end{equation}
and the summation formulas
\begin{equation}
  \sum_{k=1}^{n}\frac{(-1)^{k+1}}{k}\binom{n}{k}=\eta(n),
  \qquad
   \sum_{k=1}^{n}
     \frac{(-1)^{k+1}}{k+1}\binom{n}{k}=\frac{n}{n+1},
\end{equation}
then $a^{(d)}_{d-1}$ in Eq.~(\ref{eq:adn2}) becomes
\begin{equation}
 a^{(d)}_{d-1} 
     =\frac{d}{\ln2}\Bigl[\eta(d)- 1\Bigr],
\end{equation}
which is equal to that in Eq.~(\ref{eq:adn}).
As the second step, it will be shown that
if the statement holds for $a^{(d-1)}_{n}$ with $n=1,2,\ldots,d-2$,
then it holds for $a^{(d)}_{n}$ with $n=2,3,\ldots,d-2$,
on the basis of the recurrence relation
\begin{equation}
  a^{(d)}_n = a^{(d-1)}_n+a^{(d-1)}_{n-1},
\end{equation}
which originates from
\begin{equation}
\left(1+\epsilon\right)^d \log_2 
  \left(1+\epsilon\right) =
  \left(1+\epsilon\right) \times\left(1+\epsilon\right)^{d-1} \log_2 
  \left(1+\epsilon\right).
\end{equation}
Since this recurrence relation is satisfied by both equations,
the second step can be shown.
Accordingly, by mathematical induction starting from $d=2$ and $n=1$,
the statement that Eq.~(\ref{eq:adn2})
is equal to Eq.~(\ref{eq:adn}) for all $d$ and $n$ has been proven.
Note that Eq.~(\ref{eq:adn}) can include the case of $n=0$,
since $a^{(d)}_0=0$.

Using the coefficients $\{a^{(d)}_n\}$,
the coefficients $\{c^{(d)}_n(\lambda)\}$ can be found.
The left-hand side of Eq.~(\ref{eq:cdnDef}) can be written as
\begin{equation}
\left(\lambda^2+\epsilon\right)^d \log_2 
  \left(\lambda^2+\epsilon\right)
 =\lambda^{2d}\left(1+\frac{\epsilon}{\lambda^2}\right)^d
  \left[\log_2 \lambda^2+
  \log_2\left(1+\frac{\epsilon}{\lambda^2}\right)\right],
\label{eq:cdnDef2}
\end{equation}
while from Eq.~(\ref{eq:adnDef}),
\begin{equation}
\left(1+\frac{\epsilon}{\lambda^2}\right)^d \log_2 
  \left(1+\frac{\epsilon}{\lambda^2}\right)^d =
\sum_{n=0}^{d-1} a^{(d)}_n\left(\frac{\epsilon}{\lambda^2}\right)^n
 +O\left(\epsilon^d\right).
\label{eq:adnDef2}
\end{equation}
By substituting Eq.~(\ref{eq:adnDef2}) into Eq.~(\ref{eq:cdnDef2}),
the coefficients $\{c^{(d)}_n(\lambda)\}$ can be obtained,
as in Eq.~(\ref{eq:cdn}).

Finally, the coefficients $\{b_{n}^{(s)}\}$
of Eq.~(\ref{eq:bsnDef}) will be derived.
The coefficients $\{b_{n}^{(s)}\}$ can be found by defining $K_s$ as
\begin{equation}
 K_s\equiv \prod_{r\neq s}
  \left(1+\frac{\epsilon}{\bar{\lambda}_{ms}^2
        -\bar{\lambda}_{mr}^2}\right)^{-n_{r}}
   =\sum_{n=0}^{\infty}\, b_{n}^{(s)}\,\epsilon^n
\end{equation}
and expanding $\ln K_s$, rather than $K_s$ itself, as a Taylor series:
\begin{equation}
 \ln K_s=\sum_{n=1}^{\infty} \frac{1}{n!}\,h^{(s)}_n\,\epsilon^n,
\end{equation}
where the coefficients $\{h^{(s)}_n\}$ are given by Eq.~(\ref{eq:hsn}).
Therefore, $K_s$ can be expressed as the exponential of a Taylor series:
\begin{equation}
  K_s=\exp\left(\sum_{n=1}^{\infty}\frac{1}{n!}\, h^{(s)}_n\,\epsilon^n\right).
\label{eq:Ks}
\end{equation}
According to Fa\`a di Bruno's formula,
the exponential of a Taylor series can be expanded as a Taylor series
by the complete Bell polynomials shown in Eq.~(\ref{eq:BellPolynomial}) as
\begin{equation}
  \exp\left(\sum_{n=1}^{\infty}\frac{1}{n!}\, x_n\,\epsilon^n\right)
   =\sum_{n=0}^{\infty}\frac{1}{n!}\, 
    B_{n}\left(x_1,x_2,\ldots,x_{n}\right)\,\epsilon^n.
\label{eq:BrunoFormula}
\end{equation}
By applying this formula to Eq.~(\ref{eq:Ks}),
the coefficients $\{b_{n}^{(s)}\}$ of Eq.~(\ref{eq:bsn}) can be obtained.
Note that the complete Bell polynomials satisfy
the following formulas:
for a constant $c$ and a positive integer $m$,
\begin{align}
B_{n}\left(cx_1,c^2x_2,\ldots,c^nx_{n}\right) &=
 c^n B_{n}\left(x_1,x_2,\ldots,x_{n}\right), \\
B_{n}\bigl(0!\,m,1!\,m,\ldots,(n-1)!\,m\bigr) &= 
  n!\,\binom{m+n-1}{m-1}.
\end{align}
The first formula is valid on the basis of
the definition in Eq.~(\ref{eq:BellPolynomial}),
and the second formula can be derived
from Eq.~(\ref{eq:BrunoFormula}) because
\begin{align}
  & \sum_{n=0}^{\infty}\frac{1}{n!}\, 
    B_{n}\bigl(0!\,m,1!\,m,\ldots,(n-1)!\,m\bigr)\,\epsilon^n \notag\\
  &\quad
    =\exp\left(m\sum_{n=1}^{\infty}\frac{1}{n}\,\epsilon^n\right)
    =e^{-m\ln\left(1-\epsilon\right)}
    = \frac{1}{(1-\epsilon)^m}.
\end{align}

\section{\label{sec:limits}Limits to Projective Measurement}
Herein, Eq.~(\ref{eq:informationEx}) for $(k,l)$
is shown to be equal to Eq.~(\ref{eq:informationProj})
for $r=k$ when $\lambda=0$
and to that for $r=k+l$ when $\lambda=1$,
as expected from the definition of Eq.~(\ref{eq:measurementEx}).
When $\lambda=0$, the dangerous term $J=J_1+J_2$
given in Eqs.~(\ref{eq:dangerousJ1}) and (\ref{eq:dangerousJ2}) becomes
\begin{equation}
  \lim_{\lambda\to 0} J =
 \sum_{n=0}^{k-1}\binom{k+l-n-2}{l-1}\,(-1)^{k-n-1}\,a^{(k+l)}_n,
\end{equation}
since $c^{(d)}_n(0)=0$.
This expression can be simplified by the identity
\begin{equation}
\sum_{n=0}^{k-1}\binom{k+l-n-2}{l-1}\,(-1)^{k-n-1}\,a^{(k+l)}_n
=a^{(k)}_{k-1},
\end{equation}
which is derived from
\begin{equation}
  \frac{1}{(1+\epsilon)^l} \times\left(1+\epsilon\right)^{k+l} \log_2 
  \left(1+\epsilon\right)  
   = \left(1+\epsilon\right)^k \log_2 
  \left(1+\epsilon\right)
\end{equation}
by expanding
$1/(1+\epsilon)^l$,
$\left(1+\epsilon\right)^{k+l} \log_2 \left(1+\epsilon\right)$, and
$\left(1+\epsilon\right)^k \log_2 \left(1+\epsilon\right)$
in the Taylor series,
and comparing terms of order $\epsilon^{k-1}$ on both sides.
The dangerous term is then found to be
\begin{equation}
\lim_{\lambda\to 0} J
  =a^{(k)}_{k-1}
  =\frac{k}{\ln2}\Bigl[\eta(k)- 1\Bigr]
\end{equation}
by using Eq.~(\ref{eq:adn}).
This shows that when $\lambda=0$,
Eq.~(\ref{eq:informationEx}) becomes
\begin{equation}
I(m)=\log_2\frac{d}{k}-\frac{1}{\ln2}\Bigl[\eta(d)- \eta(k)\Bigr],
\end{equation}
which is equal to Eq.~(\ref{eq:informationProj}) for $r=k$.

On the other hand, when $\lambda=1$,
the dangerous term $J=J_1+J_2$ has apparent divergences
as in Eqs.~(\ref{eq:dangerousJ1}) and (\ref{eq:dangerousJ2}).
However, it can be calculated by
substituting $1+\epsilon$ for $\lambda^2$
and taking the limit as $\epsilon\to 0$.
Note that
the divergent terms in Eqs.~(\ref{eq:dangerousJ1})
and (\ref{eq:dangerousJ2}) should cancel each other,
since $J$ is finite.
The dangerous term is thus given by
\begin{align}
  \lim_{\lambda\to 1} J
    &=\sum_{n=0}^{l-1}\binom{k+l-n-2}{k-1}\,(-1)^{l-n-1} \notag \\
    &\qquad\times
     \left[\binom{k+l}{n}a^{(k+l-n)}_{k+l-n-1}+(k+l-n)\,a^{(k+l)}_n \right].
\end{align}
Moreover, using
\begin{equation}
  \binom{k+l}{n}a^{(k+l-n)}_{k+l-n-1}+(k+l-n)\,a^{(k+l)}_n 
   =\binom{k+l-1}{n}\,a^{(k+l)}_{k+l-1}
\end{equation}
derived from Eq.~(\ref{eq:adn}) and
\begin{equation}
\sum_{n=0}^{l-1}\,(-1)^{l-n-1} \,\binom{k+l-n-2}{k-1}\,
      \binom{k+l-1}{n}=1
\end{equation}
derived from
\begin{equation}
   \frac{1}{(1+\epsilon)^k}\times\left(1+\epsilon\right)^{k+l-1}
    =\left(1+\epsilon\right)^{l-1}
\end{equation}
by comparing terms of order $\epsilon^{l-1}$ on both sides,
it is found to be
\begin{equation}
\lim_{\lambda\to 1} J=
   a^{(k+l)}_{k+l-1}=\frac{k+l}{\ln2}\Bigl[\eta(k+l)- 1\Bigr].
\end{equation}
This shows that when $\lambda=1$,
Eq.~(\ref{eq:informationEx}) becomes
\begin{equation}
I(m)=\log_2\frac{d}{k+l}-\frac{1}{\ln2}\Bigl[\eta(d)- \eta(k+l)\Bigr],
\end{equation}
which is equal to
Eq.~(\ref{eq:informationProj}) for $r=k+l$.


\begin{thebibliography}{10}
\bibitem{Jones91}
K.~R.~W. Jones, J. Phys. A: Math. Gen. {\bf 24},  1237  (1991),
note that in the relevant equation (29)
there is a typographical error
that the first factor should be $1/d$ instead of $\Gamma(d)$.
\end{thebibliography}

\end{document}